%% file: IEEE-conference-template-062824.tex
\documentclass[conference]{IEEEtran}
\IEEEoverridecommandlockouts

\usepackage{cite}
\usepackage{amsmath,amssymb,amsfonts}
\usepackage{algorithmic}
\usepackage{graphicx}
\usepackage{textcomp}
\usepackage{xcolor}
\def\BibTeX{{\rm B\kern-.05em{\sc i\kern-.025em b}\kern-.08em
    T\kern-.1667em\lower.7ex\hbox{E}\kern-.125emX}}
\begin{document}

\title{JPEGs Just Got Snipped: Croppable Signatures Against Deepfake Images}

\author{
\IEEEauthorblockN{1\textsuperscript{st} Pericle Perazzo, 2\textsuperscript{nd} Massimiliano Mattei, 3\textsuperscript{rd} Giuseppe Anastasi,}
\IEEEauthorblockN{4\textsuperscript{th} Marco Avvenuti, 5\textsuperscript{th} Gianluca Dini, 6\textsuperscript{th} Giuseppe Lettieri, 7\textsuperscript{th} Carlo Vallati}
\IEEEauthorblockA{\textit{Dept. of Information Engineering} \\
\textit{University of Pisa}\\
Pisa, Italy \\
pericle.perazzo@unipi.it, m.mattei9@studenti.unipi.it, giuseppe.anastasi@unipi.it, \\
marco.avvenuti@unipi.it, gianluca.dini@unipi.it, giuseppe.lettieri@unipi.it, carlo.vallati@unipi.it}
}

\maketitle

\begin{abstract}
\input{00_abstract}
\end{abstract}


\input{01_introduction}
\input{02_related_work}
\input{03_proposed_method}
\input{04_jpeg_integration}
\input{05_evaluation}
\input{06_conclusions}



\section*{Acknowledgment}
This research has been partially financed by the Italian Ministry of Education and Research (MUR) within the framework of the FoReLaB project (Departments of Excellence), 
and by the European Union-NextGenerationEU within the MUR National Recovery and Resilience Plan project SERICS (PE00000014) - FF4ALL: Detection of Deep Fake Media and Life-Long Media Authentication, CUP D43C22003050001.

\bibliographystyle{IEEEtran}
\bibliography{99_bibliography}

\end{document}

%% file: 00_abstract.tex
Deepfakes are a type of synthetic media created using artificial intelligence, specifically deep learning algorithms. 
This technology can for example superimpose faces and voices onto videos, creating hyper-realistic but artificial representations. 
Deepfakes pose significant risks regarding misinformation and fake news, because they can spread false information by depicting public figures saying or doing things they never did, undermining public trust.
In this paper, we propose a method that leverages BLS signatures (Boneh, Lynn, and Shacham 2004) to implement signatures that remain valid after image cropping, but are invalidated in all the other types of manipulation, including deepfake creation. 
Our approach does not require who crops the image to know the signature private key or to be trusted in general, and it is $\mathcal{O}(1)$ in terms of signature size, making it a practical solution for scenarios where images are disseminated through web servers and cropping is the primary transformation.
Finally, we adapted the signature scheme for the JPEG standard, and we experimentally tested the size of a signed image.

%% file: 01_introduction.tex
\section{Introduction}
\label{sec:introduction}

\begin{figure}[t]
\centering
\includegraphics[trim=0 80 0 0, clip,width=1.0\linewidth]{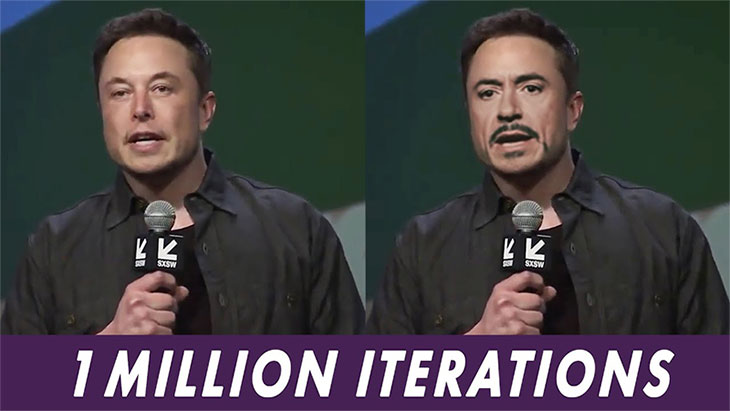}
\caption{Example of deepfake.}
\label{fig:deepfake}
\end{figure}

In recent years, the rapid advancement of artificial intelligence and machine learning technologies has given rise to the new and concerning phenomenon of \emph{deepfakes}~\cite{garg2023deepfake,khder2023artificial}. 
Deepfakes (Fig.~\ref{fig:deepfake}) are synthetic media created using deep-learning algorithms that manipulate or generate visual and audio content with remarkable realism. 
In particular, deepfake images are becoming increasingly sophisticated, making it difficult for the human eye to distinguish between real and manipulated content. 
While this technology has legitimate applications in entertainment, education, and creative industries, it also poses significant risks, particularly in the spread of misinformation, and fake news.
Deepfake images can be used to spread false information or propaganda, or to create fake evidence to undermine trust in individuals or institutions. 
The consequences can affect not only individuals but also businesses, governments, and society as a whole. 
For example, a deepfake image of a public figure in ambiguous or compromising situations could destabilize political systems, or influence political decision making by blackmail. 
Deepfakes can be used for defamation and reputation damage, making it appear that someone has engaged in inappropriate or illegal activities.
Moreover, manipulated images in journalism could erode public trust in media.

To combat the growing threat of deepfake images, there is an urgent need for robust mechanisms to authenticate digital content. 
One promising solution is the use of digital signatures, which are a cryptographic technique that ensures the integrity and the origin of a file or a piece of information in general. 
When applied to an image, a digital signatures can verify its authenticity and integrity, by confirming that the image was created by a given source and has not been altered afterwards. 
Upon downloading images from the web, browsers could display to the users trustworthy information about who created them.
Unfortunately, the usage of common signature schemes like RSA and ECDSA are made less practical by the fact that images are often edited in legitimate ways before being published on the web, for example they are cropped or reduced in resolution, etc. 
If this happens, a digital signature is of course invalidated.

In this paper, we propose a method that leverages the aggregability properties of BLS signatures~\cite{boneh2004short,perazzo2025tiny} to implement a signature scheme that remains valid after image cropping, but it is invalidated by all the other types of manipulations, including deepfake creation. 
Our approach does not require the web server to know the signature private key or to be trusted in general, and it generates only $\mathcal{O}(1)$ additional traffic volume on the web server, making it extremely practical for scenarios where image cropping is the primary transformation.
Finally, we adapt the scheme for the JPEG standard maintaining backward compatibility, and we experimentally measure the size of a signed image under varying cropping granularity levels.

The rest of the paper is structured in the following way. 
Section~\ref{sec:related_work} compares with relevant related work. 
Section~\ref{sec:proposed_method} describes in detail our proposed signature scheme. 
Section~\ref{sec:jpeg_integration} describes how it is possible to seamlessly integrate our signature scheme in the JPEG format, while maintaining the backward compatibility. 
Section~\ref{sec:evaluation} presents our experimental performance evaluation. 
Finally, the paper is concluded in Section~\ref{sec:conclusions}.

%% file: 02_related_work.tex
\section{Related Work}
\label{sec:related_work}
We divide the related work in three broad categories: international standards about image authentication (Section~\ref{ssec:standards}), authentication solutions based on zk-SNARKs (Section~\ref{ssec:zksnarks}), and homomorphic and redactable signatures (Section~\ref{ssec:homomorphic_signatures}). Our approach falls within the latter category.

\subsection{Image Signature Standards}
\label{ssec:standards}
Several standards have been proposed to ensure the authenticity and integrity of digital images. 
These standards use signed data structures (called ``manifests''), which commit to the original image data and the accompanying metadata.
They generally use common signatures like ECDSA, so if the image is cropped by an editor, he must sign another manifest that replaces or integrates the original one.
These standards assume that the image editor is trusted not to manipulate the image by creating a deepfake, which contrasts with our approach that does not require such an assumption.
Apostolopoulos et al.~\cite{apostolopoulos2006emerging} introduced the JPEG-2000 security (JPSEC) standard, which provides a framework for secure imaging, including content protection, data integrity checks, and authentication. 
More recently, the Coalition for Content Provenance and Authenticity (C2PA)~\cite{c2pa2023technical}, supported by Adobe, aims to certify the source and history of media content (images, audio, video), providing a comprehensive provenance for media assets. 
The C2PA standard is perhaps the most prominent one nowadays for media authentication, but it still relies on the assumption that the image editor is trustworthy.
The upcoming ISO JPEG Trust Standard~\cite{mo2023towards,temmermans2023towards} extends the C2PA one, and it aims to provide trustworthiness of media assets in the era of deep-learning-generated content. 
This standard also assumes that the image editor is trusted to handle the media assets appropriately.

\subsection{zk-SNARK-Based Approaches}
\label{ssec:zksnarks}
Approaches based on zero-knowledge succinct non-interactive arguments of knowledge (zk-SNARKs) offer a different method for ensuring the authenticity of digital images. 
These methods allow for more flexibility compared to our approach, because they allow for many types of image editing actions.
Not only cropping, but also scaling, compressing, and blurring.
However, they are poorly efficient in terms of proof size and verification processing time.
In this direction, Naveh and Tromer~\cite{naveh2016photoproof} proposed PhotoProof, a system for cryptographic image authentication that allows any set of permissible transformations. Kang et al.~\cite{kang2022zkimg} introduced ZK-IMG, a library for attesting to image transformations using zk-SNARKs. 
ZK-IMG allows for high-level image transformations while preserving input privacy, but it is slower and bigger than our approach.
Datta et al.~\cite{datta2024veritas} developed VerITAS, a system that uses zk-SNARKs to verify image transformations at scale. VerITAS supports large images and provides robust security, but it requires significant computational resources and bandwidth.

\subsection{Homomorphic and Redactable Signatures}
\label{ssec:homomorphic_signatures}
Homomorphic and redactable signature schemes offer another approach to ensuring the authenticity of digital images. 
These schemes typically produce logarithmic signatures, whereas our approach produces constant ones.
Johnson et al.~\cite{johnson2002homomorphic} introduced homomorphic and redactable signature schemes that allow for the construction of signatures on arbitrarily redacted submessages of the originally signed message. 
In practice, our proposed method is quite similar to the first one presented in Section 4 of~\cite{johnson2002homomorphic}, but using aggregable signatures instead of generic ones.
This allows us to dramatically save space on each edited signature with respect to Johnson et al.'s construction.
Johnson et al.~\cite{johnson2012homomorphic} extended this work to homomorphic signatures for digital photographs, allowing for the verification of image authenticity even after transformations.
With respect to our approach, these schemes are typically more efficient in the size of the original signature, as they produce a $\mathcal{O}(1)$-sized signatures, while we produce $\mathcal{O}(n)$-sized signatures, where $n$ is the size of the image to sign in blocks.
However, the size of the original signature impacts very little on the overall efficiency of the system, because each signed image is transferred only once from the image source to the server that edits it and disseminates it. 
On the other hand, the edited image is downloaded many times by the clients, so the signature on the edited image must be as small as possible. 
By using the aggragability property of the BLS signature we obtain an $\mathcal{O}(1)$ signature on the edited image, while the Johnson et al.'s one is logarithmic in size. 
We thus produce a smaller traffic on the server.
Finally, Erfurth~\cite{erfurth2024digital} developed digital signatures for authenticating compressed JPEG images. 
These signatures allow for image compression without invalidating the signature, but they do not support image cropping, thus offering a different functionality than ours.
The development of a bandwidth-efficient homomorphic signature scheme supporting both cropping and compression remains an open problem, which we plan to investigate in future work.

Our proposed method leverages short signatures to implement a redactable signature scheme specifically for JPEG cropping. 
This approach is very efficient in terms of size, making it a practical solution for scenarios where image cropping is the primary transformation.

%% file: 03_proposed_method.tex
\section{Proposed Method}
\label{sec:proposed_method}

\begin{figure}
\centering
\includegraphics[trim=0 300 370 0, clip,width=1.0\linewidth]{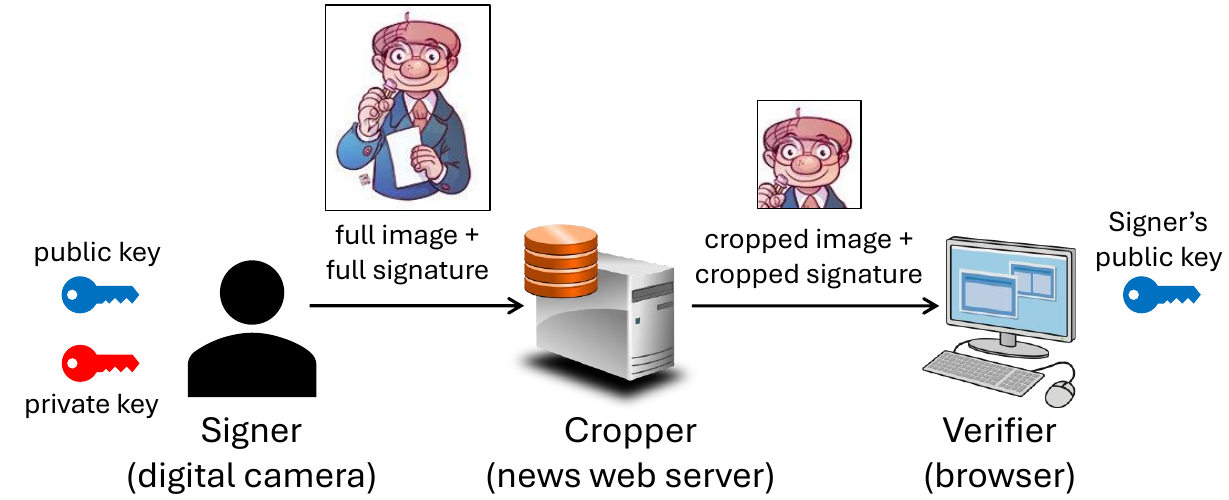}
\caption{System model.}
\label{fig:system_model}
\end{figure}

Our system model is depicted in Fig.~\ref{fig:system_model}. 
The \emph{signer} is a party that produces the original images (\emph{full images}). 
By using its private key, the signer puts a signature (\emph{full signature}) on each full image it produces before distributing it to one or more croppers. 
The signer can be for example the firmware of the digital camera that takes the pictures, as proposed by the C2PA standard~\cite{c2pa2023technical}, as well as an image elaboration software run by the a press agency.
The \emph{cropper} is a party interested in distributing images to the public, for example a news website. 
The cropper possibly crops the full image before publishing it, thus producing a \emph{cropped image}.
It also modifies the full signature by producing a \emph{cropped signature}, which authenticates the cropped image. 
To do that, the cropper does not have to know the private key of the signer, nor to have his own private key. 
The cropped signature accompanies the cropped image, for example by including it in the image file.
Finally, the \emph{verifier} is a party that wants to verify the authenticity of the cropped signature on the cropped image.
To do that, he needs to know the signer's public key, which we assume to be publicly known.
The verifier can be a web browser software, or more likely a browser extension, capable of showing authenticity information associated to the images it displays to the user, for example a green checkbox at the bottom-right corner of the image with a label ``verifiable source: digital camera X.''

Our scheme allows the image to be cropped multiple times by multiple croppers in series, as long as all of them know the full signature. 
We do not focus on this possibility in the present paper, and we assume a single cropping operation by a single cropper for the sake of simplicity.
The full signature produced by the signer can have a size linear with the image dimension, since it impacts very little on the overall efficiency of the system. 
Indeed, the full signature is transferred only once from the signer to the cropper.
On the other hand, we require the cropped signature to be extremely compact. 
This is because it is downloaded very often by verifiers, so bandwidth consumption must be minimized. 

Our proposed method is to apply the `trivial construction' for redactable signatures by Johnson et al. (Reference~\cite{johnson2002homomorphic}, Section 4), but using the aggregate BLS signature scheme by Boneh et al. (Reference~\cite{boneh2004short}, Section~3 and Section~5.1) instead of a generic signature scheme.
This allows us to drastically compact the cropped signature to an $\mathcal{O}(1)$ size. 
The BLS signature scheme is parameterized with a pairing-friendly elliptic curve, basing on which it is possible to define: (i) an additive cyclic group $\mathbb{G}_1$ of curve points, with generator $P_1$ and order $r$; (ii) an additive cyclic group $\mathbb{G}_2$ of curve points with generator $P_2$ and same order $r$; (iii) an efficient non-degenerate bilinear map $e: \mathbb{G}_1 \times \mathbb{G}_2 \longrightarrow \mathbb{G}_T$, where $\mathbb{G}_T$ is a multiplicative cyclic group with generator $e(P_1,P_2)$ and same order $r$; (iv) a hash function $H: \lbrace 0,1 \rbrace^* \longrightarrow \mathbb{G}_1$. 
We assume the above parameters are known by all entities in the system. 
We also assume that the signer owns a pair of asymmetric keys relative to a generic digital signature scheme. 
It does not have to support aggregability, so it can be any signature scheme, like ECDSA. 
We denote with $K$ the private key, with $\mathit{PK}$ the public key, with $\mathit{Sign}_K(m)$ the signature generated with the private key $K$ on a message $m$, and with $\mathit{Verify}_{\mathit{PK}}(s,m)$ the verification operation done with the public key $\mathit{PK}$ on the signature $s$ and the message $m$, which can result in a $T$ (true) or a $F$ (false) outcome. 
We assume that the signer publicly distributes his public key.
Finally, we assume that the image is divided in blocks of pixels, and these blocks are representable independently to each other. 
We denote with $w$ and $h$ respectively the width and the height of the image expressed in number of blocks, and with $x_{ij}$ (\emph{block data}) with $i \in [1, h]$ and $j \in [1, w]$ the representation of the block at coordinates $i$ and $j$. 
Therefore, the whole image is represented by the matrix: 
\begin{equation}
\begin{bmatrix} x_{11} & x_{12} & \cdots & x_{1w} \\ 
x_{21} & x_{22} & \cdots & x_{2w} \\ 
\vdots & \vdots & \ddots & \vdots \\ 
x_{h1} & x_{h2} & \cdots & x_{hw} 
\end{bmatrix}.
\end{equation} 
The requirement that block data are represented independently is needed to avoid that the single $x_{ij}$ terms change during the cropping. 
In other words, cropping corresponds simply to selecting a subset of the $x_{ij}$ terms, those inside the cropping rectangle. 
In this way, the signer can sign the single $x_{ij}$ terms and the cropper can select a subset of them without invalidating their signatures. 
To this regard, we note that some image formats are not divided in blocks (like in PNG), or even if they are, the representation of a block depends on those of precedent blocks following a differential compression encoding (like in JPEG). 
For these reasons, it is always preferable to consider the \emph{uncompressed} image representation, rather than the compressed one. 
The data blocks will be thus pieces of the uncompressed image, and the signature algorithm will be computed from them. 
This slows down a bit the signature verification, since the image must first be decompressed and then authenticated, but it is necessary for the scheme to be applicable to virtually all the image compression formats. 
We will explain an example of how to apply the scheme in the JPEG format in Section~\ref{sec:jpeg_integration}.

The proposed scheme is composed of the following three algorithms.
\begin{itemize}
\item \textbf{Sign}. 
This algorithm is run by the signer to generate a full signature on an image. 
It takes as input the private key $K$, the height $h$, the width $w$, and all the block data $x_{ij}$ with $i \in [1, h]$ and $j \in [1, w]$.
The algorithm first picks an \emph{ephemeral private key} $k$ at random in $\mathbb{Z}_r^*$, and then it computes the corresponding \emph{ephemeral public key} $\mathit{pk} = k \cdot P_2$. 
Such ephemeral private and public keys constitute a key pair of the BLS signature scheme, which the signer will use to sign only the current image. 
In this way, different images will be signed with different ephemeral keys, so that an adversary cannot replace a block of a signed image with a block of another signed image. 
For each block data $x_{ij}$ of the image, the algorithm computes a \emph{block signature} $S_{ij} = k \cdot H(i | j | x_{ij}) \in \mathbb{G}_1$, where `$|$' is the concatenation symbol. 
Each $S_{ij}$ is a BLS signature. 
Note that the ephemeral private key never signs twice the same message with BLS, because the block's indices $i$ and $j$ are signed together with the block data. 
This is important to avoid a particular attack against BLS signatures~\cite{boneh2004short}.
The output full signature will be:
\begin{equation}
S = \left( S^\prime, \forall_{1 \leq i \leq h, 1 \leq j \leq w} S_{ij} \right),
\end{equation}
where:
\begin{equation}
S^\prime = \mathit{Sign}_K(w | h | \mathit{pk}).
\end{equation}
The $S^\prime$ term authenticates the ephemeral public key with the (non-ephemeral) one. 
The $S_{ij}$ terms authenticate the various block data.
\item \textbf{Crop}. 
This algorithm is run by the cropper to generate a cropped signature, which is relative to a cropped version of the full image. 
It takes as input the full signature $S$, the block data $x_{ij}$ of the full image, and the cropping rectangle under the form of four indices: $i_{1}$ and $j_{1}$ relative to the top-left block, and $i_{2}$ and $j_{2}$ relative to the bottom-right block. 
Each block is inside the cropping rectangle if and only if its $i$ and $j$ indices fall respectively in $[i_{1}, i_{2}]$ and in $[j_{1},j_{2}]$.
Note that we do not pose any constraint on the position of the cropping rectangle inside the full image, so the image can be cropped in an arbitrary way, as long as the cropping rectangle remains aligned with the blocks.
The output cropped signature will be:
\begin{equation}
S_{i_{1},i_{2},j_{1},j_{2}} = \left( S^\prime, S^{\prime\prime}\right),
\end{equation}
where:
\begin{equation}
S^{\prime\prime} = \sum_{i_1 \le i \le i_2, j_1 \le j \le j_2} S_{ij}.
\end{equation}
The second term is a summation that corresponds to aggregating all the block signatures relative to the image blocks within the cropping rectangle. 
Note that the size of the cropped signature is constant and independent of the size of both the full and the cropped image.
This makes our solution $\mathcal{O}(1)$ in terms of authentication data exchanged between the cropper and the verifier, making it practical for the typical scenario in which the cropper is a news website and the verifier is a web browser.
\item \textbf{Verify}. 
This algorithm is run by the verifier to check the validity of a cropped signature on a particular cropped image. 
It takes as input the cropped signature $S_{i_{1},i_{2},j_{1},j_{2}}$, the blocks $x_{ij}$ of the cropped image ($i_1 \leq i \leq i_2, j_1 \leq j \leq j_2$), and the public key $\mathit{PK}$.
The algorithm first verifies the authenticity of $\mathit{pk}$ by checking that $\mathit{Verify}_{\mathit{PK}}(S^\prime,w|h|\mathit{pk}) = T$. 
Then, it verifies the authenticity of the cropped blocks by checking whether $e(S^{\prime\prime}, P_2) = \prod_{i_1 \leq i \leq i_2, j_1 \leq j \leq j_2} e(H(i | j | x_{ij}), \mathit{pk})$.
\end{itemize} 

Due to the properties of the BLS signature scheme, the verification algorithm is correct and the signature is existentially unforgeable under a chosen message attack in the random oracle model, assuming the computational Diffie-Hellman problem is hard. 
We refer to~\cite{boneh2004short} for the proof details.
From the point of view of processing capabilities, we note that while the cropper and the verifier should have plenty of resources, the signer could be somewhat limited in processing capabilities. 
Indeed, if the signer is a digital camera, it could be battery-powered, and enjoy limited processing power typical of embedded computers. 
However, BLS signatures have been proved to be fully bearable by embedded and Internet-of-Things devices by past literature~\cite{perazzo2025tiny}.

%% file: 04_jpeg_integration.tex
\section{Integration with JPEG Format}
\label{sec:jpeg_integration}

The JPEG (Joint Photographic Experts Group) compression process involves several steps: color space conversion from RGB to luminance-chrominance representation, downsampling of the crominance components, division in 8x8 blocks, Discrete Cosine Transform (DCT) on each block, quantization that reduces the resolution of high-frequency DCT components, and final Huffman coding.

The JPEG file format consists of numerous sections, each introduced by a header. 
A byte inside the header specifies which is the section type. 
Notable section types are `Start of Scan', which contains data representing all the compressed blocks, and `Comments', which contains comments, usually used to insert arbitrary metadata without compromising the image's correct display.

We employed a `Comments' section to accommodate the full signature in the full image, and the cropped signature in the cropped image. 
In this way, the signatures are embedded in the JPEG files without needing additional files. 
A JPEG viewer that does not support croppable signatures will simply ignore the `Comments' section, so that it can display the JPEG anyway.

We employed ECDSA configured with the secp160r1 standard curve for the signer key, so that $S^\prime$ is an ECDSA signature.
This curve gives approximately 80 bits of security level.
We also configured the BLS signature scheme with the BN-158 standard pairing-friendly curve.
This curve also gives approximately 80 bits of security level, so the global security level of the signature scheme is 80 bits.
Together with the signature, we also put an X.509 certificate in DER format for certifying the signer's public key.
This certificate is not part of the signature scheme described in Section~\ref{sec:proposed_method}, but it is useful for supporting a public-key infrastructure, which in turn is necessary in a real-life application.

In JPEG format, the compressed representations of the block in the `Start of Scan' section are not independent to each other. 
Rather, the constant components of each block (i.e., the ones that are independent with the spatial frequency) are represented as differences from the ones of the block on its left, while the constant components of the left-most block of each line are represented with their absolute values. 
This is a mechanism to compress even more the block representations. 
Due to this, blocks are not represented independently to each other, and so the croppable signature scheme is not applicable directly to the JPEG-compressed blocks.
We thus considered the uncompressed representation of the blocks, rather than the compressed one, using the technique explained in Section~\ref{sec:proposed_method}.

Basically, we considered the native 8x8 blocks of the JPEG format to apply our signature scheme, but we also implemented a way to sign multiples of the standard JPEG blocks.
We call `\emph{block granularity}' the integer ratio between the sides of the blocks signed with our signature scheme and the 8x8 JPEG blocks.
For example, with block granularity equal to 1 we sign every 8x8 block, with block granularity equal to 2 we sign every 16x16 block, and so on.
A coarser block granularity allows for smaller full signatures, but it also produces a degradation in the cropping precision, since the cropping rectangle must align to the blocks.

%% file: 05_evaluation.tex
\section{Evaluation}
\label{sec:evaluation}

We developed prototypes of the Sign, Crop, and Verify algorithms adapted for the JPEG format (see Section~\ref{sec:jpeg_integration}) in the C++ language, using the PBC library for the pairing-based cryptography operations, and the JED library for the encoding and decoding of JPEG files.
We tested such prototypes on real JPEG images to measure the size of produced signed images.
During the Crop algorithm execution, the full image is cropped in its top-left quarter, in such a way that the cropped image has half the width and half the height of the full one.
As a comparison, we also implemented the similar construction of Johnson et al. (Reference~\cite{johnson2002homomorphic}, Section~4).
Such a signature scheme differs to ours as it uses a standard signature scheme instead of BLS.
Therefore, the Crop algorithm cannot aggregate the block signatures, so the cropped signature size remains linear with the blocks in the cropping rectangle, that is with the size of the cropped image.
We used ECDSA configured with the secp160r1 curve to implement the scheme of Johnson et al.

\begin{figure}
\centering
\includegraphics[width=0.90\linewidth]{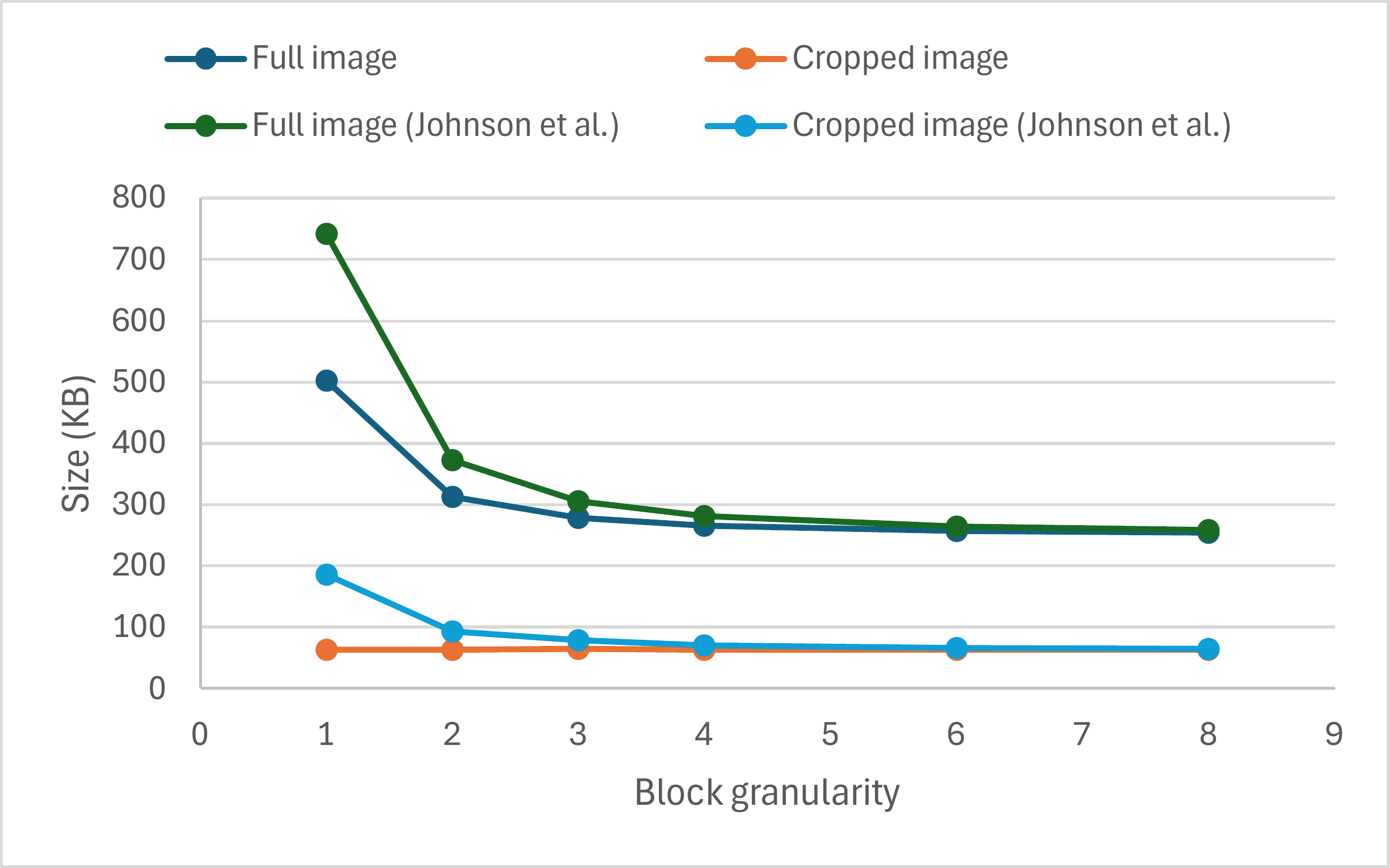}
\caption{Size of 1024x768 250 KB signed image.}
\label{fig:sizes_1024x768}
\end{figure}

Fig.~\ref{fig:sizes_1024x768} shows the size of a signed 1024x768-pixel image compressed with a medium-quality ratio, in such a way that the original unauthenticated file is 250 KB.
Such an image can be considered representative of a `medium-resolution image.'
It can be seen that the signed full image produced by our method is sensibly smaller that the one of Johnson et al.
This comes from the fact that the BLS signatures, in addition of being aggregable, are also smaller than ECDSA signatures with the same level of security.
The signed cropped image with our method is also smaller than Johnson et al.'s, this time due principally to the aggregability of the BLS signatures.
This size advantage of our method in both the full and the cropped images disappears progressively as the block granularity gets coarser, because the size of Johnson et al.'s signatures decreases.
Starting from block granularity equal to 6, corresponding to 48x48 pixel blocks, the size differences become negligible.
As we said before, a coarser block granularity produces a degradation in the cropping precision, since the cropping rectangle must align to the blocks.

\begin{figure}
\centering
\includegraphics[width=0.90\linewidth]{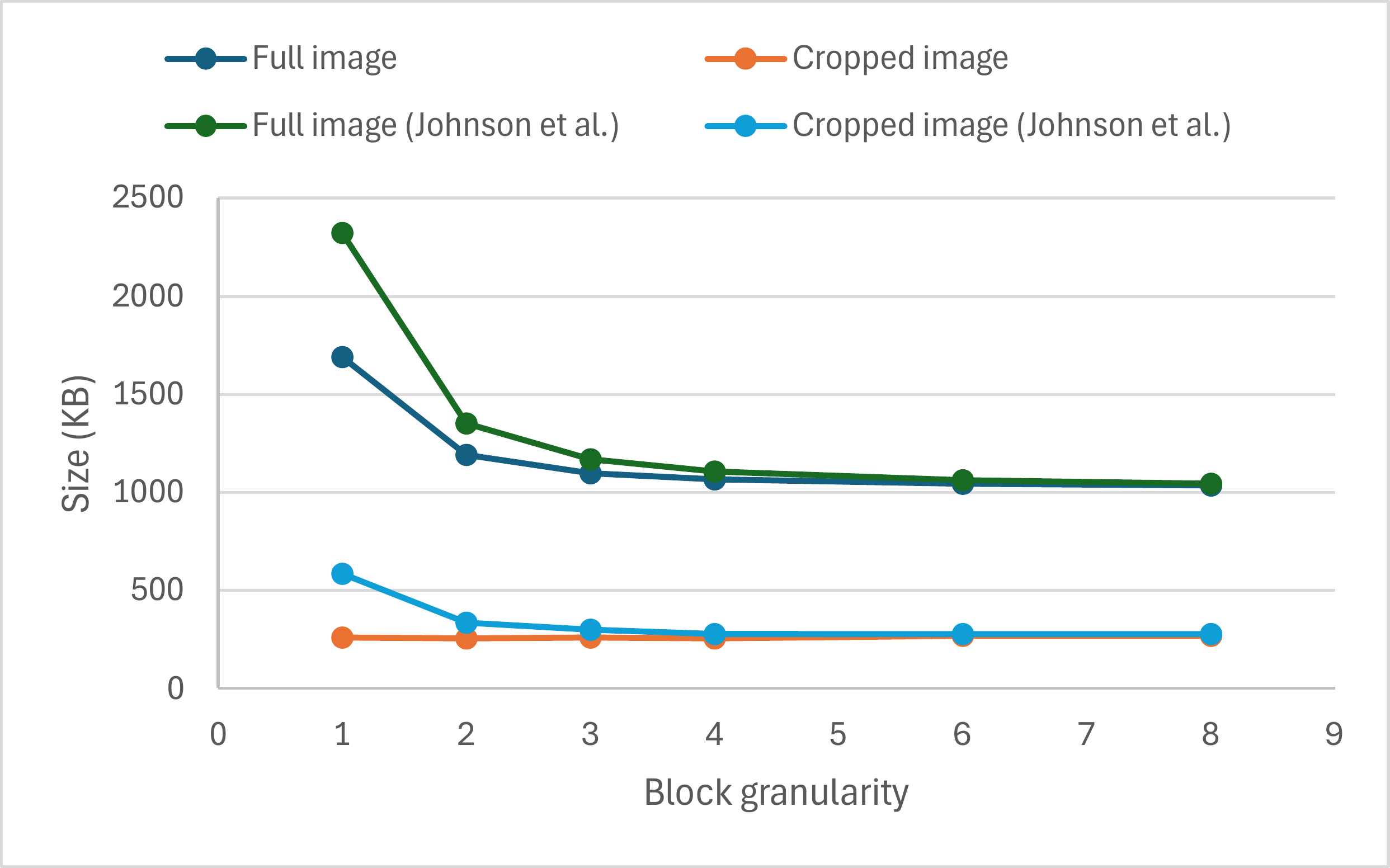}
\caption{Size of 1920x1080 1 MB signed image.}
\label{fig:sizes_1920x1080}
\end{figure}

Fig.~\ref{fig:sizes_1920x1080} shows the size of a signed 1920x1080-pixel image compressed with a medium-quality ratio, in such a way that the original unauthenticated file is 1 MB.
Such an image can be considered representative of a `medium/high-resolution image.'
All the trends of the previous graph are confirmed, in particular the size advantage of our method becomes negligible starting from block granularity equal to 6.

\begin{figure}
\centering
\includegraphics[width=0.90\linewidth]{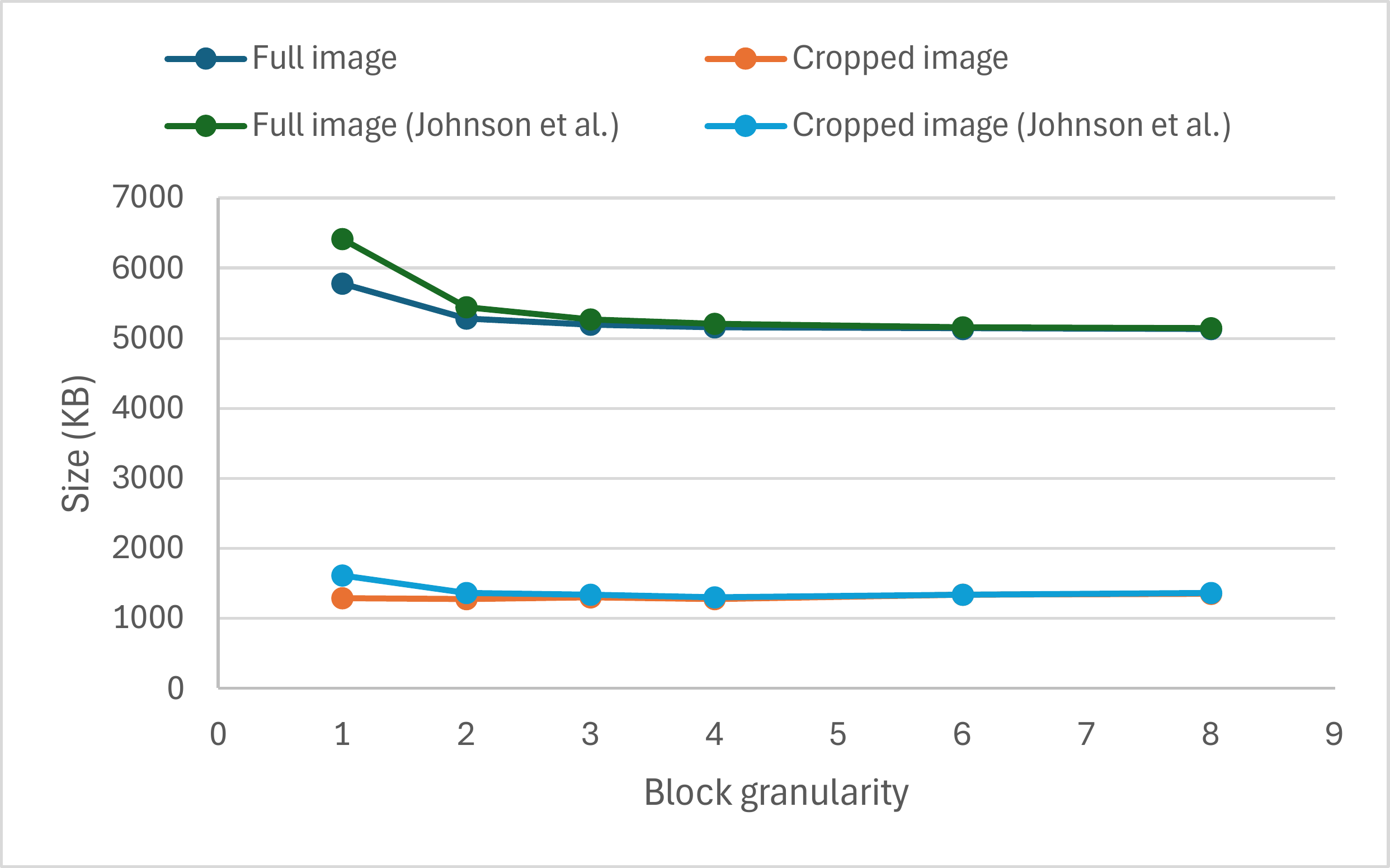}
\caption{Size of 1920x1080 5 MB signed image.}
\label{fig:sizes_1920x1080_2}
\end{figure}

Fig.~\ref{fig:sizes_1920x1080_2} shows the size of a signed 1920x1080-pixel image compressed with a high-quality ratio, in such a way that the original unauthenticated file is 5 MB.
Such an image can be considered representative of a `high-resolution image.'
We can see that the advantage of our method gets smaller with respect to previous graphs.
This is because as the resolution of the image becomes higher, the size of the signatures becomes more and more negligible with respect to the size of the image data.
However, our method continues having a non-negligible advantage up to block granularity 2, which corresponds to 16x16 pixels blocks.

%% file: 06_conclusions.tex
\section{Conclusions}
\label{sec:conclusions}

In this paper, we proposed a method that leverages the aggregability properties of BLS signatures~\cite{boneh2004short} to implement a signature scheme that remains valid after image cropping, but it is invalidated by all the other types of manipulations, including deepfake creation. 
Our approach does not require the web server to know the signature private key or to be trusted in general, and it generates only $\mathcal{O}(1)$ additional traffic volume on the web server, making it extremely practical for scenarios where image cropping is the primary transformation.
Finally, we adapted the scheme for the JPEG standard maintaining backward compatibility, and we experimentally measured the size of a signed image under varying cropping granularity levels.